\documentclass[aps,prd,reprint,superscriptaddress,showpacs,nofootinbib]{revtex4-1}

\renewcommand{\thesection}{\arabic{section}}
\renewcommand{\thesubsection}{\thesection.\arabic{subsection}}
\usepackage{hyperref}
\usepackage{amsmath}
\usepackage{amssymb}
\usepackage{graphicx}
\usepackage{color}
\usepackage{subfig}
\usepackage{bm}
\usepackage{gensymb}
\usepackage{bigints}
\usepackage[usenames,dvipsnames,svgnames,table]{xcolor}
\usepackage{pifont}
\usepackage{aas_macros}
\usepackage{orcidlink}
\usepackage{physics}

\usepackage[labelfont=bf]{caption}

\hypersetup{
    colorlinks=true,
    linkcolor=red,
    filecolor=magenta,
    urlcolor=Sepia,
    citecolor=blue,
   pdftitle={Sharelatex Example},
}
\renewcommand{\thetable}{\arabic{table}}
\numberwithin{equation}{section}

\begin{document}

\title{Equilibrium of slowly rotating polytropes in modified theories of gravity}

\author{Shaswata Chowdhury\orcidlink{.....}}
\email[E-mail: ]{shaswata@iitk.ac.in}
\affiliation{Department of Physics, Indian Institute of Technology, Kanpur 208016, India}
\author{Pritam Banerjee\orcidlink{.....}}
\email[E-mail: ]{bpritam@iitk.ac.in}
\affiliation{Department of Physics, Indian Institute of Technology, Kanpur 208016, India}
\author{Aneta Wojnar\orcidlink{0000-0002-1545-1483}}
\thanks{Corresponding author}
\email[E-mail: ]{awojnar@ucm.es}
\affiliation{Departamento de F\'isica Te\'orica, Universidad Complutense de Madrid, E-28040, Madrid, Spain}

\begin{abstract}
A general formalism to find the density profile of a slowly rotating stellar object in modified gravity is presented. We derive a generic Lane-Emden equation and its analytical solution for a wide class of modified theories of gravity. 
\end{abstract}

\maketitle


\section{Introduction}

A polytropic equation of state (EoS) turns out to be a useful approximation to describe matter properties in substellar and stellar objects, as well as neutron stars. It has a simple form
\begin{equation}
P=K\rho^{1+\frac{1}{n}},
\label{Poly}
\end{equation}
relating stellar density $\rho$ to the pressure $P$, where $K$ and $n$ are polytropic parameters or functions, taking different
expressions and values being dependent upon the class of stellar objects we are considering. Because of that, it
allows one to analyze, often analytically, a given astrophysical object in modified theories of gravity \cite{sotiriou2010f,2011PhR...509..167C,nojiri2011unified,nojiri2017modified} before applying a more complex approach, with more realistic microphysics. However, even with such a simple form, many sophisticated processes can be hidden in this EoS. The most important one is the electron degeneracy, crucial in modelling some layers of the Sun \cite{saltas2019obtaining,saltas2022searching} and other Main Sequence stars \cite{davis2012modified,sakstein2013stellar}, low-mass stars \cite{bildsten1997lithium,ushomirsky1998light,koyama2015astrophysical}, brown dwarfs, and giant exoplanets \cite{burrows1993science} as well as white dwarfs \cite{2015JCAP...05..045D,saito2015modified,bertolami2016white,saltas2018white,astashenok2022maximal,astashenok2022chandrasekhar}. Another improvement which can also be incorporated into microphysics modelling and then rewritten in the polytropic form are strongly coupled plasma \cite{stevenson1991search}, finite gas temperatures with phase transition points between metallic hydrogen and molecular state \cite{auddy2016analytic}. Moreover, a merger of the third-order finite strain Birch-Murgnagham equation of state \cite{birch1947finite} with Thomas-Fermi-Dirac one \cite{thomas1927calculation,fermi1927statistical,dirac1930note,feynman1949equations,salpeter1967theoretical} turns out to be also approximated by the polytropic EoS \cite{seager2007mass}, which is suitable to describe matter behaviour in cold low-mass spheres such as terrestrial planets. 

The set of following equations: polytropic EoS \eqref{Poly}, Poisson
equation ($G$ is the Newton's gravitational constant, with $U$ being the gravitational potential)
\begin{equation}
\nabla^2U = -4\pi G \rho,
\label{Poisson0}
\end{equation}
together with the equation of hydrostatic equilibrium in Newtonian gravity, both considered in the spherical-symmetric spacetime 
\begin{align}
\frac{1}{r^2}\frac{d}{d r}\Big(r^2 \frac{d U}{d r}\Big) &=-4\pi G \rho
\label{PoissonNew}\\
\frac{d P}{d r} &= \rho\frac{d U}{d r}
\label{mechanical}
\end{align}
can be rewritten into 
the Lane-Emden equation (LEE)
\begin{equation}
\frac{1}{\xi^2}\frac{d}{d \xi}\Big(\xi^2 \frac{d \theta}{d \xi}\Big) =-\theta^n,
\label{modLEEn}
\end{equation}
where $\theta$ is a function of $\xi$, satisfying the boundary conditions $\theta(0)=1$, $\theta'(0)=0$, with $'$ denoting derivative with respect to $\xi$.
To do so, one needs to introduce the dimensionless variables $\theta$ and $\xi$, such that
\begin{equation}
\rho=\rho_c \theta^n, ~~~~~ r=r_c\xi ~~~~~{\rm with}~~~~~ r_c^2=\frac{K(n+1)\rho_c^{(\frac{1}{n}-1)}}{4\pi G}.
\label{nondimensional0}
\end{equation}
where $\rho_c$ denotes the central density.
The solution of the LEE with a particular value of the polytropic index $n$ and polytropic constant $K$ provides the total stellar mass $M$, stellar radius $R$, and the density profile \eqref{nondimensional0}, pressure \eqref{Poly}, temperature $T$ as well as the core quantities $\rho_c$ and $T_c$:
\begin{eqnarray}
 M&=&4\pi r_c^3\rho_c\omega_n,\;\;\;
 R=\gamma_n\left(\frac{K}{G}\right)^\frac{n}{3-n}M^\frac{1-n}{n-3},\\
 \rho_c&=&\delta_n\left(\frac{3M}{4\pi R^3}\right) \label{rho0s},\;\;\;
 T=T_c\theta=K \frac{m_H\mu }{k_B}\rho_c^\frac{1}{n}\theta,\label{rhoc}
\end{eqnarray}
 where $k_B$ is Boltzmann's constant, $m_H$ is the mass of Hydrogen atom, and $\mu$ being the mean molecular weight while
\begin{eqnarray}
 \omega_n&=&-\xi_1^2\frac{d\theta}{d\xi}\Big|_{\xi=\xi_1},\label{omega}\\
  \gamma_n&=&(4\pi)^\frac{1}{n-3}(n+1)^\frac{n}{3-n}\omega_n^\frac{n-1}{3-n}\xi_1,\label{gamma}\\
 \delta_n&=&-\frac{\xi_1}{3\frac{d\theta}{d\xi}|_{\xi=\xi_1}} . \label{delta}
\end{eqnarray}
For more properties, see \cite{horedt2004polytropes}.

On the other hand, modified theories of gravity (MG) often introduce additional terms to Poisson equation \cite{banados2010eddington,koyama2015astrophysical,Olmo:2019flu,toniato2020palatini,olmo2021parameterized}, which we can write in a generic form as:  
\begin{equation}
\frac{1}{r^2}\frac{d}{d r}\Big(r^2 \frac{d U}{d r}\Big) = -4\pi G \rho + LV_{mod0}(r),
\label{Poisson1}
\end{equation}
where the modification term $LV_{mod0}(r)$ is a general function being different for different classes of modified gravity theories. As before, introducing to it the polytropic EoS \eqref{Poly}, the pressure balance equation \eqref{mechanical} along with the dimensionless quantities \eqref{nondimensional0}, we can write down
the modified Lane-Emden equation (MLEE)
\begin{equation}
\frac{1}{\xi^2}\frac{d}{d \xi}\Big(\xi^2 \frac{d \theta}{d \xi}\Big) =-\theta^n +  g_{mod0}(\xi)
\label{modLEEn}
\end{equation}
where the extra term appearing in the above
\begin{equation}\label{gmod}
g_{mod0}=\frac{LV_{mod0}}{4\pi G \rho_c}
\end{equation}
is a dimensionless term induced by a given MG.

So far, most of the stellar and substellar objects have been studied in MG with some form of polytropic EoS, however in the spherical-symmetric spacetime (for review, see~\cite{Olmo:2019flu,CANTATA:2021ktz,Wojnar:2022txk}).
In order to obtain limiting masses, such as, for instance, the Chandrasekhar mass-limit of white dwarfs \cite{chandrasekhar1935highly,saltas2018white,jain2016white,banerjee2017constraints,wojnar2021white,belfaqih2021white,2022PhRvD.105b4028S,2022PhLB..82736942K,2021ApJ...909...65K,2019ApJ...884...95C}, the minimum Main Sequence mass~\cite{sakstein2015hydrogen,sakstein2015testing,crisostomi2019vainshtein,Olmo:2019qsj}, minimum mass for deuterium burning~\cite{rosyadi2019brown}, Jeans~\cite{capozziello2012jeans} and opacity mass~\cite{wojnar2021jupiter}, the authors were using the considered EoS. To constrain models of gravity with the use of seismic data from stars \cite{saltas2019obtaining,saltas2022searching}, and rocky planets \cite{Kozak:2021ghd,Kozak:2021zva,Kozak:2021fjy}, polytropic EoS was also applied to describe at least some of the object's layers. In a similar manner, to obtain light elements' abundances \cite{Wojnar:2020frr}, the polytrope was also adopted, the same as in the evolutionary phases of various astrophysical objects \cite{Wojnar:2020txr,chowdhury2021modified,Guerrero:2021fnz, straight2020modified,Gomes:2022sft,chowdhury2022revisiting,Benito:2021ywe,Kozak:2022hdy,Kalita:2022zki,Wojnar:2022ttc,Kalita:2022trq}.

Although the astrophysical objects rotate, rotating polytropes \cite{chandrasekhar1933equilibrium,monaghan1965structure,1968ApJ...154..999K,cook1994rapidly,kong2015exact,Chowdhury:2021idh,ballot2009gravity} (for the second order approximation, see \cite{occhionero1967rotationally}) have not been studied widely in the non-relativistic framework of MG. Several studies of stellar rotation in MG in the relativistic framework are found in the literature. Rotating neutron stars have been studied in the context of various MG theories using polytropic equation of state such as scalar-tensor theories \cite{Doneva2013PRD}, dilatonic Einstein–Gauss–Bonnet theory \cite{Kleihaus2016PRD}, and Rastall’s gravity \cite{da2021rapidly}. In the relativistic context \cite{komatsu1989rapidly,paschalidis2017rotating} one usually uses the numerical approach, which is even more inevitable in the case of modifications introduced to the Tolman-Oppenheimer-Volkoff equation by MG (for review, see \cite{Olmo:2019flu} and references therein). Because of this poorly studied branch, we are going to look for an analytic density profile given for a wide class of MG in the non-relativistic regime. Interestingly, MG's effects in the stellar interior are prominent even in the non-relativistic limit.

The paper is organized as follows. In section \ref{analytic}, we provide a form of the modified Lane-Emden equation for a rotating polytrope together with its analytical solution. Those are given for a wide class of modified theories of gravity, which satisfies three specific conditions as demonstrated in the further part of the section. Section \ref{ExampleSphericalb-H} is devoted to specific examples of MG, which allows us to put forward a corollary on the conditions to be satisfied by a modified Poisson equation, such that the presented formalism is valid. We draw our conclusions in section \ref{conl}.


\section{analytic derivation for the density profile of a slowly rotating stellar object in modified theories of gravity }\label{analytic}

In this section, we present the analytical formalism to incorporate slow rotation in any modified theories of gravity in general. We would be specifically following the approach of \cite{chandrasekhar1933equilibrium}. 
Therefore, we will consider the rotation of the object to be along the Z-axis of the 3D Cartesian coordinate system, with the uniform angular speed denoted by $\omega$. In polar coordinates \{$r, \mu, \phi$\}, where $r$ is the radial coordinate, and $\mu(=\cos\vartheta)$, $\phi$ being the angular coordinates, the equations of mechanical equilibrium are as follows:
\begin{align}
\frac{\partial P}{\partial r} &= \rho\frac{\partial V}{\partial r} + \rho\omega^2r(1-\mu^2)~,\label{mechanical1}\\
\frac{\partial P}{\partial \mu} &= \rho\frac{\partial V}{\partial \mu} - \rho\omega^2r^2\mu
\label{mechanical2}
\end{align}
with $\phi$ being neglected on account of axial symmetry, and abiding by the convention of \cite{chandrasekhar1933equilibrium}, $V$ is chosen to be the negative of the gravitational potential energy. Although no additional terms due to modified gravity theories appear explicitly in Eq. (\ref{mechanical1}) and \eqref{mechanical2}, the potential $V$ inherently captures the effect of MG. However, in modified gravity theories the Poisson equation gets modified, as can be seen, for example, from \cite{banados2010eddington}, \cite{Hernandez-Arboleda:2022rim},
and \cite{banerjee2021constraining}
\begin{equation}
\nabla^2V = -4\pi G \rho + LV_{mod}(r,\mu).
\label{Poisson}
\end{equation}
In the above Eq. (\ref{Poisson}), we refer to the modification term $LV_{mod}(r,\mu)$ as a general function, without mentioning its actual form; the form is going to be different for different classes of modified gravity theories. The $\mu$ dependence in the modification term is induced by the rotation, i.e., in the absence of rotation the modification term will solely depend upon the radial coordinate $r$\footnote{For example in beyond-Horndeski, the density perturbation to the FRW metric is in general spherically symmetric in the absence of rotation.}, leading to Eq.\eqref{Poisson1}. Therefore, regarding the matter description, we take the total pressure $P$ inside a rotating stellar object to be related to its density $\rho=\rho(r,\mu)$ by means of a polytropic relation:
\begin{equation}
P(r,\mu)=K\rho^{1+\frac{1}{n}}.
\label{Polytrope}
\end{equation} 
We emphasize the fact that the pressure and density are functions of both the radial as well as the angular coordinate in presence of rotation-induced asymmetry. Analogously to the non-rotating case, we can
introduce the dimensionless variables $\Theta$ and $\xi$, such that
\begin{equation}
\rho=\rho_c \Theta^n, ~~~ r=r_c\xi ~~~{\rm with}~~~ r_c^2=\frac{K(n+1)\rho_c^{(\frac{1}{n}-1)}}{4\pi G} 
\label{nondimensional}
\end{equation}
where $\Theta$ is a function of both $\xi$ and $\mu$. It should be noted that Eq.(\ref{nondimensional}) is similar to Eq.(\ref{nondimensional0}), other than $\theta(\xi)$ being replaced by $\Theta(\xi,\mu)$. 

To go further, let us propose the following:\\[2ex]

\textbf{Proposition:} Let $\theta=\theta(\xi)$ be a solution of the modified Lane-Emden equation \eqref{modLEEn} of the non-rotating polytrope \eqref{Poly} and $\Theta=\Theta(\xi,\mu)$ of the rotating one \eqref{Polytrope} in the polar coordinates. Then, the Lane-Emden equation for a rotating polytrope with the uniform angular speed $\omega$ in modified gravity is given by
\begin{align}
&\frac{1}{\xi^2}\frac{\partial}{\partial \xi}\Big(\xi^2 \frac{\partial \Theta}{\partial \xi}\Big) + \frac{1}{\xi^2}\frac{\partial}{\partial \mu}\Big((1-\mu^2) \frac{\partial \Theta}{\partial \mu}\Big) \nonumber
\\
&= v + g_{mod}(\xi,\mu)-\Theta^n
\label{modLEE}
\end{align}
where $v=\omega^2/2\pi G \rho_c$ is a dimensionless parameter, which is a measure of the outward centrifugal force compared to the self-gravity of the rotating polytrope, while $g_{mod}(\xi,\mu)=LV_{mod}/4\pi G \rho_c$ is the dimensionless modification term depending on a given theory of gravity in general.
\\[2ex]
\textit{Proof:} \\[2ex]
The Poisson equation \eqref{Poisson} in polar coordinates takes the form
\begin{align}
&\frac{1}{r^2}\frac{\partial}{\partial r}\Big(r^2 \frac{\partial V}{\partial r}\Big) + \frac{1}{r^2}\frac{\partial}{\partial \mu}\Big((1-\mu^2) \frac{\partial V}{\partial \mu}\Big)\nonumber\\
&=-4\pi G \rho +   LV_{mod}(r,\mu).
\label{PoissonReduced}
\end{align}
Using the mechanical equilibrium equations (\ref{mechanical1}), (\ref{mechanical2}), along with the polytropic EoS \eqref{Polytrope} and Eq.(\ref{nondimensional}), the Poisson equation (\ref{PoissonReduced}) reduces to the modified Lane-Emden equation (MLEE)
\begin{align*}
&\frac{1}{\xi^2}\frac{\partial}{\partial \xi}\Big(\xi^2 \frac{\partial \Theta}{\partial \xi}\Big) + \frac{1}{\xi^2}\frac{\partial}{\partial \mu}\Big((1-\mu^2) \frac{\partial \Theta}{\partial \mu}\Big)\\\
&= v + g_{mod}(\xi,\mu) -\Theta^n 
\end{align*}
 Therefore, we have demonstrated the general form of the MLEE for the rotating polytrope \eqref{modLEE}.
\rightline{Q.E.D.}\\[2ex]

Note that for a given central density, the parameter $v$, being the measure of the strength of rotation, will be the expansion parameter for certain functions and solutions in the sequel. Hereafter, such an exact form of the modified Lane-Emden equation in the case of rotation allows now to study rotating objects. Usually, one solves this equation numerically. However, we may also try to get the analytical solution - this is particularly useful because one can track the effects of modified gravity and easily compare it with the Newtonian case \cite{chandrasekhar1933equilibrium}. Moreover, having an analytic solution allows us to distinguish the modifications introduced by a given theory of gravity from the other effects, like the ones coming from, e.g. microphysics or other processes which happen in the stellar and substellar interiors. Therefore, let us propose the following theorem:\\[2ex]

\textbf{Theorem:} If $g_{mod}(\xi,\mu)$ can be expanded in terms of the Legendre functions $P_l(\mu)$'s as
\begin{align}\label{TheoremEq1}
g_{mod}(\xi,\mu) &= g_{mod0}(\xi) \\
&+v\Bigg\{\bar{\tilde{g}}_{mod}(\xi)P_0(\mu) + \sum_{j=1}^{\infty} \bar{\bar{\tilde{g}}}_{modj}(\xi)P_j(\mu)\Bigg\},\nonumber
\end{align}
where $g_{mod0}(\xi)$ is the non-rotating part, with $\bar{\tilde{g}}_{mod}(\xi)$, and $\bar{\bar{\tilde{g}}}_{modj}(\xi)$ being the rotation induced ones, then the solution $\Theta$ of the modified Lane-Emden equation in presence of slow rotation is
\begin{align}
\Theta(\xi,\mu)&=\theta(\xi) \\
&+ v\Big[\psi_0(\xi) + A_2\psi_2(\xi)P_2(\mu)\Big]\nonumber,
\end{align}
where $\psi_0$ and $\psi_2$ satisfy the following equations:
\begin{equation}
\frac{1}{\xi^2}\frac{d}{d \xi}\Big(\xi^2 \frac{d \psi_0}{d \xi}\Big) = - n\theta^{n-1}\psi_0 + 1 + \bar{\tilde{g}}_{mod}(\xi),
\end{equation}
\begin{equation}
\frac{1}{\xi^2}\frac{d}{d \xi}\Big(\xi^2 \frac{d \psi_2}{d \xi}\Big) = \Big(\frac{6}{\xi^2}-n\theta^{n-1}\Big)\psi_2 + \frac{\bar{\bar{\tilde{g}}}_{mod2}(\xi)}{A_2},
\end{equation}
where
\begin{equation}
A_2= - \frac{5}{6}\frac{\xi_1^2}{[3\psi_2(\xi_1)+\xi_1 \psi_2^{'}(\xi_1)]},
\label{A2}
\end{equation}
with $\xi_1$ being the first zero of $\theta(\xi)$ while $'$ denotes derivative with respect to $\xi$.
\\[2ex]
\textit{Proof:} \\[2ex]
Let us make the following choice for the modification term $g_{mod}$
\begin{equation}
g_{mod}(\xi,\mu) = g_{mod0}(\xi) +v\tilde{g}_{mod}(\xi,\mu),
\label{Property1}
\end{equation}
where $g_{mod0}(\xi)$ is the standard modification term coming from modified gravity theories in the non-rotating scenario Eq.\eqref{modLEEn}, while $\tilde{g}_{mod}(\xi,\mu)$ is the correction term appearing in the modified gravity theories when rotation is taken into consideration. The above expansion will enable us to extract out terms in orders of $v$, in the subsequent calculations, as we will see shortly.

In order to find a solution to Eq.(\ref{modLEE}), we assume the following form for $\Theta$
\begin{equation}
\Theta(\xi,\mu) = \theta(\xi) + v\Psi(\xi,\mu) +v^2\Phi(\xi,\mu)
\label{ThetaExpansion}
\end{equation}
where $\theta$ is the non-rotating solution, with $\Psi$ and $\Phi$ being the rotation induced correction terms. We are considering slow rotation where the effects arising from $\omega^4$ can be neglected. Therefore, we consistently work only up to the first order in $v$. Putting Eq.(\ref{ThetaExpansion}) in Eq.(\ref{modLEE}), the $O(v^0)$ gives back Eq.(\ref{modLEEn}) as expected, while $O(v)$ gives the following equation
\begin{align}
&\frac{1}{\xi^2}\frac{\partial}{\partial \xi}\Big(\xi^2 \frac{\partial \Psi}{\partial \xi}\Big) + \frac{1}{\xi^2}\frac{\partial}{\partial \mu}\Big((1-\mu^2) \frac{\partial \Psi}{\partial \mu}\Big)\nonumber\\
&=-n\theta^{n-1}\Psi + 1 + \tilde{g}_{mod}(\xi,\mu)
\label{PsiEqn}
\end{align}
Now, for a given stellar object (i.e. fixing $n$) and a theory of modified gravity in non-rotating scenario (i.e. knowing the form of $g_{mod0}(\xi)$), we know the solution $\theta(\xi)$ from Eq.(\ref{modLEEn}). Therefore, all we need to find is the solution $\Psi(\xi,\mu)$ from Eq.(\ref{PsiEqn}), in order to obtain the complete solution $\Theta$. For that, we assume the following form for $\Psi$
\begin{equation}
\Psi(\xi,\mu) = \psi_0(\xi) + \sum_{j=1}^{\infty} A_j \psi_j(\xi)P_j(\mu)
\label{PsiExpansion}
\end{equation}
where $A_j$'s are arbitrary constants and $P_j(\mu)$ corresponds to Legendre function of index $j$, satisfying the Legendre differential equation
\begin{equation}
\frac{\partial}{\partial \mu}\Big((1-\mu^2)\frac{\partial P_j}{\partial \mu}\Big) + j(j+1)P_j = 0.
\label{LegendreEqn}
\end{equation}
Substituting Eq.(\ref{PsiExpansion}) in Eq.(\ref{PsiEqn}) and using Eq.(\ref{LegendreEqn}) we get
\begin{align}
&0=\Big[\frac{1}{\xi^2}\frac{d}{d \xi}\Big(\xi^2 \frac{d \psi_0}{d \xi}\Big)
+ n\theta^{n-1}\psi_0 - 1\Big] + \tilde{g}_{mod}(\xi,\mu)\nonumber\\
&+ \sum_{j=1}^{\infty} A_j\Big[\frac{1}{\xi^2}\frac{d}{d \xi}\Big(\xi^2 \frac{d \psi_j}{d \xi}\Big) - \Big(\frac{j(j+1)}{\xi^2}-n\theta^{n-1}\Big)\psi_j\Big]P_j(\mu) 
\label{PsiEqnExpand}
\end{align}
 From the above equation, it is clearly seen that the modification term $\tilde{g}_{mod}$ couples to the independent terms associated to the linearly independent Legendre functions, and thus forbids their complete extraction and equating them to zero. One possible way of averting the situation is by having $\tilde{g}_{mod}$ of this particular form:
\begin{equation}
\tilde{g}_{mod}(\xi,\mu) = \bar{\tilde{g}}_{mod}(\xi) + \sum_{j=1}^{\infty} \bar{\bar{\tilde{g}}}_{modj}(\xi)P_j(\mu)
\label{Property2}
\end{equation}
Upon using Eq.(\ref{Property2}) in Eq.(\ref{PsiEqnExpand}) and equating coefficients of the linearly independent Legendre functions, we get
\begin{equation}
\frac{1}{\xi^2}\frac{d}{d \xi}\Big(\xi^2 \frac{d \psi_0}{d \xi}\Big) = - n\theta^{n-1}\psi_0 + 1 + \bar{\tilde{g}}_{mod}(\xi)
\label{psi0}
\end{equation}
\begin{equation}
\frac{1}{\xi^2}\frac{d}{d \xi}\Big(\xi^2 \frac{d \psi_j}{d \xi}\Big) = \Big(\frac{j(j+1)}{\xi^2}-n\theta^{n-1}\Big)\psi_j + \frac{\bar{\bar{\tilde{g}}}_{modj}(\xi)}{A_j}
\label{psij}
\end{equation}
Now, for a given theory of modified gravity, upon knowing the explicit forms of $\bar{\tilde{g}}_{mod}(\xi)$ and $\bar{\bar{\tilde{g}}}_{modj}(\xi)$, one can solve for $\psi_0$ and $\psi_j$ from the above Eq.(\ref{psi0}) and Eq.(\ref{psij}). At this point, it is important to mention that solving Eq.(\ref{psij}) for $\psi_j$, seems improbable due to the existence of $A_j$, which is not known beforehand. However, for such modified gravity theories, in which the modification term $\bar{\bar{\tilde{g}}}_{modj}(\xi)$ inherently carries a factor of $A_j$, one can solve for $\psi_j$. In sequel, we will show with examples that it does happen for certain classes of modified gravity theories.

Assuming for the time being that there exist modified gravity theories for which one can solve for $\psi_0$ and $\psi_j$'s using the above set of equations (\ref{psi0}) and (\ref{psij}), we are still left with determining the unknown arbitrary constants $A_j$'s. For that we will first determine $V$ in terms of $A_j$'s using Poisson equation and equations of mechanical equilibrium. Now, at the stellar surface, this $V$ must correspond to a physically viable generic form of the potential exterior (say, $V_{ext}$) to the object. This necessitates equating $V$ with $V_{ext}$ as well as the radial-derivative of $V$ with that of $V_{ext}$. Upon doing this, we will obtain the coefficients $A_j$'s in terms of the known solutions. We explicitly develop the formalism as follows.

Poisson equation (\ref{PoissonReduced}) in $\xi,\mu$ variables takes the form (to the first order in $v$)
\begin{align}\label{PoissonReducedDimless}
&\frac{1}{\xi^2}\frac{\partial}{\partial \xi}\Big(\xi^2 \frac{\partial V}{\partial \xi}\Big) + \frac{1}{\xi^2}\frac{\partial}{\partial \mu}\Big((1-\mu^2) \frac{\partial V}{\partial \mu}\Big)\\
&=-(n+1)K\rho_c^{\frac{1}{n}}\Big[\theta^n + n\theta^{n-1}v\Big\{\psi_0 + \sum_{j=1}^{\infty}A_j\psi_j(\xi)P_j(\mu)\Big\}\Big]\nonumber
\end{align}
In order to solve for $V$ in the above equation, we develop $V$ in the form (to the first order in $v$)
\begin{equation}
V=U(\xi)+v\Big\{V_0(\xi)+\sum_{j=1}^{\infty}V_j(\xi)P_j(\mu)\Big\},
\label{Vexpansion}
\end{equation}
where $U$ is the modified gravity potential of the non-rotating configuration. Upon using Eq.(\ref{Vexpansion}) in Eq.(\ref{PoissonReducedDimless}), and then equating $O(v^0)$ component and coefficients of $P_j(\mu)$ in $O(v)$ component, we get
\begin{equation}
\frac{1}{\xi^2}\frac{d}{d \xi}\Big(\xi^2 \frac{d U}{d \xi}\Big) = -R\Big\{\theta^n - g_{mod0}(\xi)\Big\}
\label{Ueqn}
\end{equation}
\begin{equation}
\frac{1}{\xi^2}\frac{d}{d \xi}\Big(\xi^2 \frac{d V_0}{d \xi}\Big) = -R\Big\{n\theta^{n-1}\psi_0 - \bar{\tilde{g}}_{mod}(\xi)\Big\}
\label{V0eqn}
\end{equation}
\begin{equation}
\frac{1}{\xi^2}\frac{d}{d \xi}\Big(\xi^2 \frac{d V_j}{d \xi}\Big) - j(j+1)V_j = -R\Big\{n\theta^{n-1}A_j\psi_j - \bar{\bar{\tilde{g}}}_{modj}(\xi)\Big\},
\label{Vjeqn}
\end{equation}
where $R:=(n+1)K\rho_c^\frac{1}{n}$.
Although, it appears that one might not achieve analytic solutions to the above set of equations, we will show that it is not true; we do get analytic forms of the potential functions $U$, $V_0$, and $V_j$'s, after a little bit of algebra. At this point, it is convenient to mention that obtaining the analytic forms of the aforementioned functions is an integral part of our entire formalism; had we not been able to obtain the same, our entire analytic approach would have broken down at this stage.\\

Using Eq.(\ref{modLEEn}) in Eq.(\ref{Ueqn}) we obtain
\begin{equation}
\frac{1}{\xi^2}\frac{d}{d \xi}\Big(\xi^2 \frac{d U}{d \xi}\Big) = \frac{R}{\xi^2}\frac{d}{d \xi}\Big(\xi^2 \frac{d \theta}{d \xi}\Big)
\end{equation}
whereby we deduce
\begin{equation}
U=R\theta + const
\end{equation}
Although this analytic form looks exactly the same in case of standard Newtonian gravity (see \cite{chandrasekhar1933equilibrium}), the difference lies in the fact that here $\theta$, being solution to Eq.(\ref{modLEEn}), is implicitly carrying the information of the modified gravity theory under consideration. Such an information is also included in $R$, as it depends on $\rho_c$, which is given by \eqref{rhoc} and \eqref{delta}.

Using Eq.(\ref{psi0}) in Eq.(\ref{V0eqn}) and Eq.(\ref{psij}) in Eq.(\ref{Vjeqn}) we obtain
\begin{equation}
\frac{1}{\xi^2}\frac{d}{d \xi}\Big(\xi^2 \frac{d V_0}{d \xi}\Big) = R\Big[\frac{1}{\xi^2}\frac{d}{d \xi}\Big(\xi^2 \frac{d \theta}{d \xi}\Big) -1\Big],
\end{equation}
\begin{align}
&\frac{1}{\xi^2}\frac{d}{d \xi}\Big(\xi^2 \frac{d V_j}{d \xi}\Big) -j(j+1)V_j\\
&= RA_j\Big[\frac{1}{\xi^2}\frac{d}{d \xi}\Big(\xi^2 \frac{d \psi_j}{d \xi}\Big) - \frac{j(j+1)}{\xi^2}\psi_j \Big]\nonumber,
\end{align}
whereby we deduce
\begin{equation}
V_0=R\Big(\psi_0-\frac{1}{6}\xi^2\Big) + {\rm const},
\end{equation}
\begin{equation}
V_j=R\Big(A_j\psi_j + B_j\xi^j\Big) + {\rm const},
\end{equation}
where $B_j$ are arbitrary constants appearing from the regular solution of the Eq.(\ref{Vjeqn}). After rearranging terms we get
\begin{equation}
V=R\Big[\Theta+v\Big\{\sum_{j=1}^{\infty}B_j\xi^jP_j(\mu)-\frac{1}{6}\xi^2\Big\}\Big],
\label{Vform}
\end{equation}
where once again we mention that although this analytic form looks exactly the same in case of standard Newtonian gravity (see \cite{chandrasekhar1933equilibrium}), the information of modified gravity theory is carried implicitly by the solution $\Theta$, as pointed out above.

After converting the first relation of Eq.(\ref{mechanical2}) into its dimensionless form (by using Eq.(\ref{nondimensional}) and Eq.\ref{Polytrope}), we substitute Eq.(\ref{Vform}) for $V$ in the same. Equating coefficients of $P_j(\mu)$ we obtain
\begin{equation}
B_j=0~~\forall~~j\neq 2;~~~~~B_2=\frac{1}{6}.
\label{Bvalues}
\end{equation}
Thus we have
\begin{equation}
V=R\Big[\Theta - \frac{1}{6}v\Big(\xi^2-P_2(\mu)\xi^2\Big)\Big]+{\rm const}
\label{VanalyticForm}
\end{equation}
where the arbitrary constants $A_j$'s contained within $\Theta$ are still unknown. The $A_j$'s now get determined by implementing the condition of continuity of $V$ as well as its radial-derivative at the stellar surface. For that we assume a particular form of the gravitational potential exterior to the stellar surface 
\begin{equation}
V_{ext}=R\Big[\frac{C_0}{\xi}+v\sum_{j=1}^{\infty}\frac{C_j}{\xi^{j+1}}P_j(\mu)\Big] + {\rm const},
\label{VextForm}
\end{equation} 
where $C_i,\;i=\{0,1,...\}$ are arbitrary constants.
Now, since a rotating stellar object is oblate, it does not have a single well defined radius defining the stellar surface. On the other hand, analytic values of the stellar radii, corresponding to different angular coordinates (let us call them angular stellar radii), can only be obtained if one has the complete solution $\Theta$. Unfortunately, without knowing $A_j$'s one cannot obtain $\Theta$ and thus the angular stellar radii. Thus with the spirit of analytic formalism, we choose the first zero $\xi_1$ of the Emden's function corresponding to non-rotating modified gravity scenario as the point where we enforce the continuity of the potential and its radial-derivative in the rotating scenario i.e.,
\begin{equation}
V|_{\xi_1}=V_{ext}|_{\xi_1},~~~~~\frac{\partial V}{\partial \xi}\Big|_{\xi_1}=\frac{\partial V_{ext}}{\partial \xi}\Big|_{\xi_1}
\label{MatchingCondition}
\end{equation}
Since we are considering slow rotation, where the degree of oblateness is not high, this approximation is well justified. Now, implementing the aforementioned condition Eq.(\ref{MatchingCondition}) we get,
\begin{equation}
A_j=0~~\forall~~j\neq 2;~~~~~A_2=-\frac{5}{6}\frac{\xi_1^2}{[3\psi_2(\xi_1)+\xi_1\psi_2^{'}(\xi_1)]}.
\label{Avalues}
\end{equation}
Again, the form of $A_2$ is the same as in the standard Newtonian gravity, although the effects of modified gravity theory is encoded in $\xi_1$ and $\psi_2$ implicitly. Having obtained the $A_j$'s we now write the complete analytic solution of the MLEE,
\begin{align*}
\Theta(\xi,\mu)&=\theta(\xi) \\
&+ v\Big[\psi_0(\xi) - \frac{5}{6}\frac{\xi_1^2}{[3\psi_2(\xi_1)+\xi_1\psi_2^{'}(\xi_1)]}\psi_2(\xi)P_2(\mu)\Big]\nonumber
\end{align*}
where $\psi_0$ and $\psi_2$ are solutions to the following equations, obtained from Eq.(\ref{psi0}) and Eq.(\ref{psij}), respectively
\begin{equation*}
\frac{1}{\xi^2}\frac{d}{d \xi}\Big(\xi^2 \frac{d \psi_0}{d \xi}\Big) = - n\theta^{n-1}\psi_0 + 1 + \bar{\tilde{g}}_{mod}(\xi)
\end{equation*}
\begin{equation}
\frac{1}{\xi^2}\frac{d}{d \xi}\Big(\xi^2 \frac{d \psi_2}{d \xi}\Big) = \Big(\frac{6}{\xi^2}-n\theta^{n-1}\Big)\psi_2 + \frac{\bar{\bar{\tilde{g}}}_{mod2}(\xi)}{A_2},
\label{psi2}
\end{equation}
\rightline{Q.E.D.}\\[2ex]

Recalling the discussion after Eq.(\ref{psij}), we see from 
Eq.(\ref{psi2}), that in order to solve for $\psi_2$ one needs to know $A_2$, which itself depends upon the solution $\psi_2$. Therefore, unless the term $(\bar{\bar{\tilde{g}}}_{mod2}(\xi)/A_2)$ in entirety is independent of $A_2$, one cannot solve for the complete solution in this analytic formalism. We, therefore, enlist the three conditions, which the generic correction term $g_{mod}$ due to modified gravity theories should satisfy, in order to obtain the complete solution in this particular analytic formalism\footnote{We note that in case this third property does not get satisfied for certain class of modified gravity theories, one can employ a self-iterating numerical scheme to obtain a solution for such coupled equations. Developing such a scheme is nevertheless a daunting task, and we are not aware of such an attempt till now.}:
\begin{enumerate}
	\item $g_{mod}$ can be expanded as in Eq.(\ref{Property1}) to the first order in $v$.
	\item The $O(v)$ correction term $\tilde{g}_{mod}$ can be expanded as in Eq.(\ref{Property2}) in terms of Legendre functions.
	\item The $\bar{\bar{\tilde{g}}}_{modj}$ term should contain $A_j$, thus making $\Big(\bar{\bar{\tilde{g}}}_{modj}/A_j\Big)$ term of Eq.(\ref{psij}) entirely independent of $A_j$
\end{enumerate}

In the next section \ref{ExampleSphericalb-H} we show that the aforementioned three conditions are indeed satisfied for a vast majority of modified gravity theories.

\section{On the generic $g_{mod}$ term}
\label{ExampleSphericalb-H}
In this section, we prove that one can carry out the above analytical formalism in most of the modified gravity theories in the literature. For that we begin by reviewing how the Poisson equation gets modified in three of the most popular and working models of modified gravity theories.
\begin{itemize}
\item In generalized beyond-Horndeski theories \cite{horndeski1974second,gleyzes2015new,gleyzes2015exploring} the Poisson equation takes the form (because of the partial breaking of Vainshtein mechanism \cite{vainshtein1972problem, kobayashi2015breaking, Crisostomi2018PRD}) \cite{saltas2018white, koyama2015astrophysical}:
	\begin{equation}
	\nabla^2 V \sim -\frac{\kappa}{2}\Big(\rho+\frac{\Upsilon}{4}\nabla^2(r^2\rho)\Big)
	\label{PoissonBH}
	\end{equation}
	where $\Upsilon$ is the modified gravity parameter for the beyond-Horndeski class of theories. 
	\item In Palatini $f(R)$ gravity, the Poisson equation reads (\cite{toniato2020palatini})
	\begin{equation}
	\nabla^2 V \sim -\frac{\kappa}{2}\Big(\rho+2\beta\nabla^2\rho\Big)
	\label{PoissonPalatini}
	\end{equation}
	where $\kappa=8\pi G$ and $\beta$ is a constant\footnote{$\beta$ is of dimension ${\rm [L]^2}$, where [L] corresponds to length dimension.} associated to the $O(R^2)$ term of the function $f(R)$, with $R$ being the Palatini-Ricci scalar. The constant $\beta$ thus parametrizes this particular class of modified gravity theories. 
	\item In Eddington-inspired Born–Infeld (EiBI) gravity, the Poisson equation reads (\cite{banados2010eddington}, \cite{jimenez2018born},\cite{Banerjee2022EiBI})
	\begin{equation}
	\nabla^2 V \sim -\frac{\kappa}{2}\Big(\rho+\frac{\epsilon}{2}\nabla^2\rho\Big)
	\label{PoissonEiBI}
	\end{equation}
	where $\epsilon=1/M_{BI}$, with $M_{BI}$ being the Born-Infeld mass. $\epsilon$ is the modified gravity parameter for this class of theories. 
\end{itemize}

From the above forms of the modified Poisson equation in the different classes of modified gravity theories, we can write the modification term in general as
\begin{equation}
LV_{mod}(r,\mu) = k_1\nabla^2\Big(\bar{\alpha}(r)\rho\Big)
\label{LVmodform}
\end{equation}
where $k_1$ represents the overall constant comprising of fundamental constant $G$, numerical factors and the associated modified gravity parameter. The term $\bar{\alpha}(r)$ is in general a radial function, which for example takes up the constant value $1$ for the Palatini $f(R)$ and EiBI, while it is $r^2$ for generalized beyond-Horndeski. Converting the above equation into its non-dimensional form and multiplying it with a factor of $1/4\pi G \rho_c$, we obtain
\begin{equation}
\tilde{k}_1\nabla_{\xi}^2\Big(\alpha(\xi)\Theta^n\Big)
\label{gmodform}
\end{equation}
where $\nabla_{\xi}^2$ is the dimensionless Laplacian, and $\tilde{k}_1$ is the overall constant appearing upfront, which in general depends on $r_c$, and the modified gravity parameter. The function $\alpha(\xi)$ is the non-dimensional version of $\bar{\alpha}(r)$. To put things into perspective, let us mention that Eq.(\ref{LVmodform}) corresponds to the $LV_{mod}$ term in Eq.(\ref{Poisson}), while Eq.(\ref{gmodform}) represents $g_{mod}$ term in Eq.(\ref{modLEE}). Thus, we will investigate whether this generic $g_{mod}$ term satisfies the three conditions mentioned in the section \ref{analytic}, which are required for our analytic formalism to go through.\\

Expressing Eq.(\ref{gmodform}) explicitly we have
\begin{align}\label{gmodLaplacian}
&g_{mod}(\xi,\mu)=\tilde{k}_1\nabla_{\xi}^2\Big(\alpha(\xi)\Theta^n\Big)\\ 
&= \tilde{k}_1\Big[\frac{1}{\xi^2}\frac{\partial}{\partial \xi}\Big(\xi^2 \frac{\partial (\alpha\Theta^n)}{\partial \xi}\Big) + \frac{1}{\xi^2}\frac{\partial}{\partial \mu}\Big((1-\mu^2) \frac{\partial (\alpha\Theta^n)}{\partial \mu}\Big)\Big].\nonumber
\end{align}
Then using $\Theta^n = \theta^n + vn\theta^{n-1}\Psi$ (to the first order in $v$ from Eq.(\ref{ThetaExpansion})) in Eq.(\ref{gmodLaplacian}) we obtain
\begin{equation}
g_{mod}(\xi,\mu)= \Xi(\xi) + v\Big\{\Xi_0(\xi) + \sum_{j=1}^{\infty}A_j\Xi_j(\xi)P_j(\mu)\Big\}
\label{gmodGeneralExpression}
\end{equation}
where
\begin{equation}
\begin{split}
\Xi(\xi)&=\frac{\tilde{k}_1}{\xi}\theta^{n-2}\Big\{n(n-1)\xi \alpha {\theta'}^2 + \theta^2(2\alpha' + \xi \alpha '') \\
&+ n\theta\Big(2(\alpha + \xi \alpha ')\theta ' +\xi\alpha\theta ''\Big)\Big\}
\end{split}
\end{equation}
\begin{equation}
\begin{split}
\Xi_0(\xi) & = \frac{\tilde{k}_1}{\xi}n\theta^{n-3}\Big\{(n-2)(n-1)\xi \alpha \psi_0 {\theta'}^2\\
&+ (n-1)\theta\Big(2 \theta ' (\xi \psi_0\alpha' + \alpha(\psi_0+\xi\psi_0')) + \xi \alpha\psi_0\theta ''\Big) \\
 &  + \theta^2\Big(2(\alpha+\xi\alpha')\psi_0' + \psi_0(2\alpha'+\xi\alpha'')+\xi\alpha\psi_0''\Big)\Big\}
\end{split}
\end{equation}
\begin{equation}
\begin{split}
\Xi_j(\xi) & = \frac{\tilde{k}_1}{\xi}\Big[n\theta^{n-3}\Big\{(n-2)(n-1)\xi \alpha \psi_j {\theta'}^2 \\
&+ (n-1)\theta\Big(2 \theta ' (\xi \psi_j\alpha' + \alpha(\psi_j+\xi\psi_j')) + \xi \alpha\psi_j\theta ''\Big) \\
&  + \theta^2\Big(2(\alpha+\xi\alpha')\psi_j' + \psi_j(2\alpha'+\xi\alpha'')+\xi\alpha\psi_j''\Big)\Big\} \\
&- \frac{1}{\xi}j(j+1) n\alpha\theta^{n-1}\psi_j\Big]
\end{split}
\end{equation}
where $'$ denotes first order derivative with respect to $\xi$ and $''$ denotes second order one. Comparing Eq.(\ref{gmodGeneralExpression}) with Eq.(\ref{Property1}) and Eq.(\ref{Property2}) we identify that $\Xi$ corresponds to $g_{mod0}$ term, while $\Xi_0$ and $A_j\Xi_j$ correspond to $\bar{\tilde{g}}_{mod}$ and $\bar{\bar{\tilde{g}}}_{modj}$ terms respectively. Therefore, we see that all the three conditions enlisted in the previous section get satisfied for these broad classes of modified gravity theories and hence one can use our analytical formalism in these theories.

At this stage, it is convenient to propose our hypothesis:\\

\textbf{Corollary}
{\it In general, our analytical formalism can be used in any modified gravity theories, where the correction term of the corresponding Poisson equation contains density, its higher order radial-derivatives, or its Laplacian.}

\section{Conclusion}\label{conl}
In this work, we have presented a general formalism to find the density profile of a slowly rotating stellar object in presence of modified gravity. By adapting the formalism given in \cite{chandrasekhar1933equilibrium}, we demonstrated a generic approach to incorporate modified gravity effects. We have shown three conditions that the additional modified gravity term arising in the Poisson equation needs to satisfy in order to abide by our formalism. Firstly, the modified term is required to be expanded into a summation between non-rotating and rotating counterparts.  Secondly, the rotating part should be further expanded in series involving Legendre functions. Finally, the coefficients appearing with the Legendre functions need to explicitly involve density, its derivative terms or its Laplacian. We have undertaken three well known theories of modified gravity, i.e., scalar-tensor theories beyond-Horndeski, Palatini $f(R)$ gravity, and Eddington-inspired Born-Infeld gravity which are shown to satisfy all these three conditions. 

As already mentioned, this work is focused on slow rotation, which is necessary to set the matching conditions, Eq.(\ref{MatchingCondition}), at the first zero of a spherically symmetric non-rotating configuration $\theta$. While it is valid for a slowly rotating polytrope, it prevents one from finding a solution for fast rotation. To overcome this limitation, one must follow a semi-analytic approach, where fast rotation is achieved using multiple small increments of stellar rotation. The matching condition is reused iteratively at the first zero of the last updated rotating Emden's function $\Theta$. To this end, we draw the reader's attention to the fact that the parameter $v$, being a measure of the ratio of outward centrifugal force and self-gravity, can incorporate fast rotation for higher central density and yet be small enough to neglect $\mathcal{O}(v^2)$ corrections. Let us also emphasize that the central density can increase when a theory parameter increases in the modification term - then, in modified gravity theories, the same $v$ can correspond to larger rotation $\omega$ due to increase in $\rho_c$. It is so because the parameter $v$ we expand the solution $\Theta$ about includes the central density $\rho_c^{-1}$, lowering its value. Because of that fact, this approximation breaks down for a specific large value of $v$ in Newtonian physics, while in modified gravity one can still consider more rapidly rotating objects. 

To summarize, this work enables us to analytically obtain the density profile of a slowly rotating star by elegantly utilizing its axial symmetry. This is a stepping stone for further studies in modeling of rotating stellar and planetary objects in presence of modified gravity. The provided formalism will allow us to find a complete solution of a further specified modified Lane-Emden equation. We will present the results on the overall rotating density profile in future work.

\section*{Acknowledgements}
 AW acknowledges financial support from MICINN (Spain) {\it Ayuda Juan de la Cierva - incorporac\'ion} 2020 No. IJC2020-044751-I.

\bibliographystyle{apsrev4-1}
\bibliography{biblio}

\end{document}